\documentclass[aps,pre,twocolumn,superscriptaddress,amsmath,amssymb]{revtex4-1}
\usepackage{graphicx,bm}

\begin{document}

\title{Analysis of the high dimensional naming game with committed minorities}

\author{William Pickering}
\affiliation{Department of Mathematical Sciences, Rensselaer Polytechnic Institute, 110 8th Street, Troy, New York 12180, USA}
\affiliation{Network Science and Technology Center, Rensselaer Polytechnic Institute, 110 8th Street, Troy, New York 12180, USA}
\author{Boleslaw K. Szymanski}
\affiliation{Department of Computational Intelligence, Wroclaw University of Technology, 50-370 Wroclaw, Poland}
\affiliation{Network Science and Technology Center, Rensselaer Polytechnic Institute, 110 8th Street, Troy, New York 12180, USA}
\author{Chjan Lim}
\affiliation{Department of Mathematical Sciences, Rensselaer Polytechnic Institute, 110 8th Street, Troy, New York 12180, USA}
\affiliation{Network Science and Technology Center, Rensselaer Polytechnic Institute, 110 8th Street, Troy, New York 12180, USA}

\date{\today}
\begin{abstract}
The naming game has become an archetype for linguistic evolution and mathematical social behavioral analysis. In the model presented here, there are $N$ individuals and $K$ words. Our contribution is developing a robust method that handles the case when $K = O(N)$. The initial condition plays a crucial role in the ordering of the system. We find that the system with high Shannon entropy has a higher consensus time and a lower critical fraction of zealots compared to low-entropy states. We also show that the critical number of committed agents decreases with the number of opinions and grows with the community size for each word. These results complement earlier conclusions that diversity of opinion is essential for evolution; without it, the system stagnates in the status quo [S. A. Marvel et al., Phys. Rev. Lett. 109, 118702 (2012)]. In contrast, our results suggest that committed minorities can more easily conquer highly diverse systems, showing them to be inherently unstable.
\end{abstract}


\maketitle

\section{Introduction}

The study of sociology and political science by means of mathematical and physical principles have been increasingly popular recently \cite{sen,galam,galam2}. One of the fundamental problems in this area is the spread of opinion via social influence often represented by the voter model, in which individuals adopt the states of their neighbors  \cite{liggett, liggett2,clifford,castellano,sen}. Other related models of social influence include social impact theory \cite{sen,nowak}, threshold models \cite{granovetter}, and the naming game \cite{steels,baronchelli3,baronchelli2,xie,zhang,waagen,verma}. Here, we use the naming game as the archetype for social influence, and investigate the role of high opinion diversity on social systems \cite{waagen,mistry,xie,galam3,xie2}.

We chose the naming game because, unlike other models, it can account for several historical precedents in which the majority opinion was overtaken by a committed minority (e.g., the suffragette movement in the early 20th century, and the adoption of the American civil-rights in 1960s \cite{xie}). Such processes are known in sociology under the term minority influence \cite{moscovici1969}. When the committed minority fraction of the population is small, their opinion will still be suppressed by an existing majority opinion \cite{marvel}. Yet, when this fraction exceeds a modest tipping point value \cite{xie,couzin}, the minority opinion will spread rapidly. 

Here we aim to establish that the naming game model can also account for dynamics of opinion spread in extreme initial conditions. Our motivating historical precedents are the dynamics of postrevolution opinion struggle. Often before revolution happens, the government identifies and suppresses the leading opposition minorities that are on the verge of achieving tipping fraction of support (e.g., Islamists before Iranian revolution of 1979 or Muslim Brothers before Egyptian revolution of 2011), so the revolution is conducted by a motley of opposition movements with different ideologies united only by opposition to the government. After the revolution, the winners remove suppression of such minorities, allowing them to quickly win the majority of the population in agreement with the naming game model. However, the case of the Russian revolution of February 1917 was different. The revolt was spontaneous, disorganized, and after they won, no dominant minority exceeding the tipping point fraction of the population emerged as in the previous examples. Yet, in the midst of the disorder and dissent, a small Bolshevik party grasped the power and support of uncommitted individuals by November 1917, because their leader Lenin correctly diagnosed that the power laid on the streets. Here we study the case resembling such situations in the context of naming game, when there are committed minorities of multiple opinions. In \cite{waagen}, Waagen \textit{et al.} show that in such a case, a stalemate of opinion can more easily occur, in which no decision is reached. Similar transitions may occur without committed agents by modifying the strategies of individuals with multiple opinions, which leads to additional equilibrium states  \cite{baronchelli5,thompson}. In contrast, we identify the new set of conditions for this case under which the loss of stability of a social system occurs. Under these conditions, instead of stagnation with no decision, a rapid change occurs in which a small minority quickly spreads their opinion to the uncommitted subpopulation. In addition, we show that in the presence of committed minorities, as opinion diversity of the uncommitted subpopulation increases, the size of the committed minority needed to the turn the uncommitted to the minority opinion decreases. In extreme cases, this critical committed minority is invariant of the system size. This suggests that too much dissent between individuals makes them susceptible to even a few zealots.

To gain these insights into the dynamics of social systems, we solve the critical problem of complexity for the naming game. For $K$ opinions, the system of ordinary differential equations (ODEs) that describe relative population sizes has $2^{K}-1$ equations, which is numerically and analytically difficult to study \cite{waagen}. Furthermore, if the number of opinions also becomes infinite with $N$, then these ODE methods fail. By applying more robust methods of analysis, we solve the problem of exponential complexity, and by doing so, demonstrate the potential of solving other highly complex systems by these means. 

The format of the article is as follows. Section \ref{sec:NG} describes the details of the naming game model. The solutions that we provide are given in terms of the dominant eigenvalues of the system, which are found in Sec. \ref{sec:convergence}. Once we know the long time behavior of the model, we calculate the time to eliminate a word from the system, the expected number of words over time, and the time to reach consensus in Sec. \ref{sec:no_zealots}. Then we introduce committed minorities in the system (defined in Sec. \ref{sec:NG}), and analyze the tipping points in Sec. \ref{sec:zealots}.

\section{Characteristics of the K-word listener-only naming game}\label{sec:NG}
Here we will describe the naming game model in detail together with the notation that we will use throughout this paper. We use the listener only naming game given in Ref. \cite{baronchelli4}, which is a slight variation of the model in Ref. \cite{baronchelli3}. In the model there are $K$ words (opinions), which we call $A_1,A_2,\ldots,A_K$. In social contexts, the words in the naming game are associated with  opinions, beliefs, or political allegiances. There are $N$ individuals, each with a word list, which is a set of words. The individuals update their word lists as they change their opinions in response to messages from others. We also assume that any individual may speak to any other individual. This means that the social network is a complete graph, which is a common assumption \cite{xie,waagen,verma,maity,steels,baronchelli3,baronchelli2}, although other networks have also been considered \cite{lu,lu2}. 

It has been observed that dynamics of the naming game on real world networks are qualitatively similar to complete graph results \cite{waagen,dall'asta}. In Ref. \cite{zhang2}, it has been shown with agreement with numerical simulations that the naming game behavior is consistent over Erd\''{o}s-R\'{e}nyi (ER) networks with varying average degree. This consistency between ER networks and the complete graph is also true for the voter model \cite{pickering2}. However, when the network exhibits a strong community structure, additional equilibrium states can emerge with different communities hold different opinions \cite{lu}. 

Time is discretized so that one interaction of individuals takes place within a time step. In one step, an individual is chosen uniformly at random to be the speaker and another is chosen uniformly at random as the listener. Let $W_s$ and $W_l$ be the word lists of the speaker and the listener, respectively. The speaker chooses a random word $A_s$ in its word list to transmit to the listener. If none of the two is committed, they update their word lists according to the following rules:

\begin{enumerate}
\item If $A_s\not\in W_l$, then $W_l\rightarrow W_l\cup \{A_s\}$.
\item If $A_s\in W_l$, then $W_l\rightarrow \{A_s\},\{A_s\}$ \cite{baronchelli4}.
\end{enumerate}

In brief, if the listener does not have the spoken word in its list, then it adds it to its list. If the listener has the spoken word in its list, the listener reduces its list to the spoken word. Only the listener changes its word list as a result of an interaction, which is a slight modification of the original naming game. It has been shown in Ref. \cite{baronchelli4} that the naming game and the listener-only variant have qualitatively similar behavior in the complete graph case.

In addition to these rules, we also may include committed agents (also known as zealots) in the system. A zealot never changes their word list and adopts only a single opinion. We consider two cases when these committed minorities are present. We first consider the case when there are $n'$ zealots of one word. Then we consider the case when there are $n'$ zealots for each word. We show that there are similar rates of convergence for both cases in Sec. \ref{sec:convergence}. The critical fraction of committed agents is the value of $n'/N$ that yields a phase transition in the system. When this fraction of zealots is below this critical value, the opinion of the committed minorities will be suppressed by the majority. When the committed fraction is above the critical value, the minority opinion overcomes the majority. We are also interested in the time until all individuals have the same opinion, which we define as the consensus time. This is discussed in detail in Sec. \ref{sec:zealots}.

We initialize the system by assigning a word list to each individual. For simplicity, we assign one of the $K$ words to an individual. That is, no individual initially has a mixed word list. However, the system will quickly saturate itself with word lists with length $2$ or more \cite{baronchelli4,baronchelli2}. Also, in our mathematical analysis, we assume that there is equal representation for each word. In Sec. \ref{sec:entropy}, we use the Shannon Entropy to numerically consider the case in which there is unequal representation in the initial distribution of words.

\subsection{Shannon entropy}\label{sec:entropy}

Entropy in the naming game is a measurement of the amount of disagreement and conflict in the system. The Shannon entropy in particular measures the uncertainty of a random variable, such as a message \cite{shannon}. In the naming game, there clearly are messages that are transmitted from person to person and the entropy of these messages also has a clear social meaning. If the system has high Shannon entropy, then a listener has a significant probability of hearing a diverse range of opinions. This also means that there is greater competition among the opinions for dominance in the system. There is more dissent, disorder, and disagreement in high-entropy systems. In a low-entropy system, the listener is more likely to hear the same message consistently. Low-entropy systems have more consistency in the messages that are transmitted, so there is more agreement within the population. These systems are predictable, ordered, and united.

In one step of the naming game, a single opinion is spoken to the listener. This spoken word is a random variable that takes values in $A_1,\ldots,A_K$ with probability distribution that depends on the macrostate of the system. Let the probability of speaking $A_s$ be $P_s$. By definition, the Shannon entropy of the system is given by 
\begin{equation}
\mathcal{H}=-\sum_{s=1}^K P_s\ln P_s\label{entropy_formula}.
\end{equation}
We take the natural logarithm in Eq. \eqref{entropy_formula} for convenience. To find the probability $P_s$ of speaking each word, we let $|W_j|$ be the length of the word list corresponding to individual $j$ for $j\in\{1,\ldots,N\}$. With this, $P_s$ can be expressed as
\begin{equation}
P_s=\frac{1}{N}\sum_{j=1}^N \frac{\mathbf{1}_{W_j}(A_s)}{|W_j|}\label{P_def}
\end{equation}
where $\mathbf{1}_{W_j}$ is the indicator function defined by, 
\begin{equation}
\mathbf{1}_{W_l}(A_s)=
\begin{cases}
1 & A_s\in W_l \\
0 & A_s\not\in W_l.
\end{cases}
\end{equation}
So, given the word lists of every node in the network, we use Eqs. \eqref{P_def} and \eqref{entropy_formula} to calculate the entropy of the system. With this definition, we aim to demonstrate numerically and analytically the following entropy principles: 
\begin{enumerate}
\item The consensus time is expected to increase as $\mathcal{H}$ increases.
\item The critical fraction of committed agents is expected to decrease as $\mathcal{H}$ decreases. 
\end{enumerate}

\begin{figure}[h!]
\includegraphics[scale=.4]{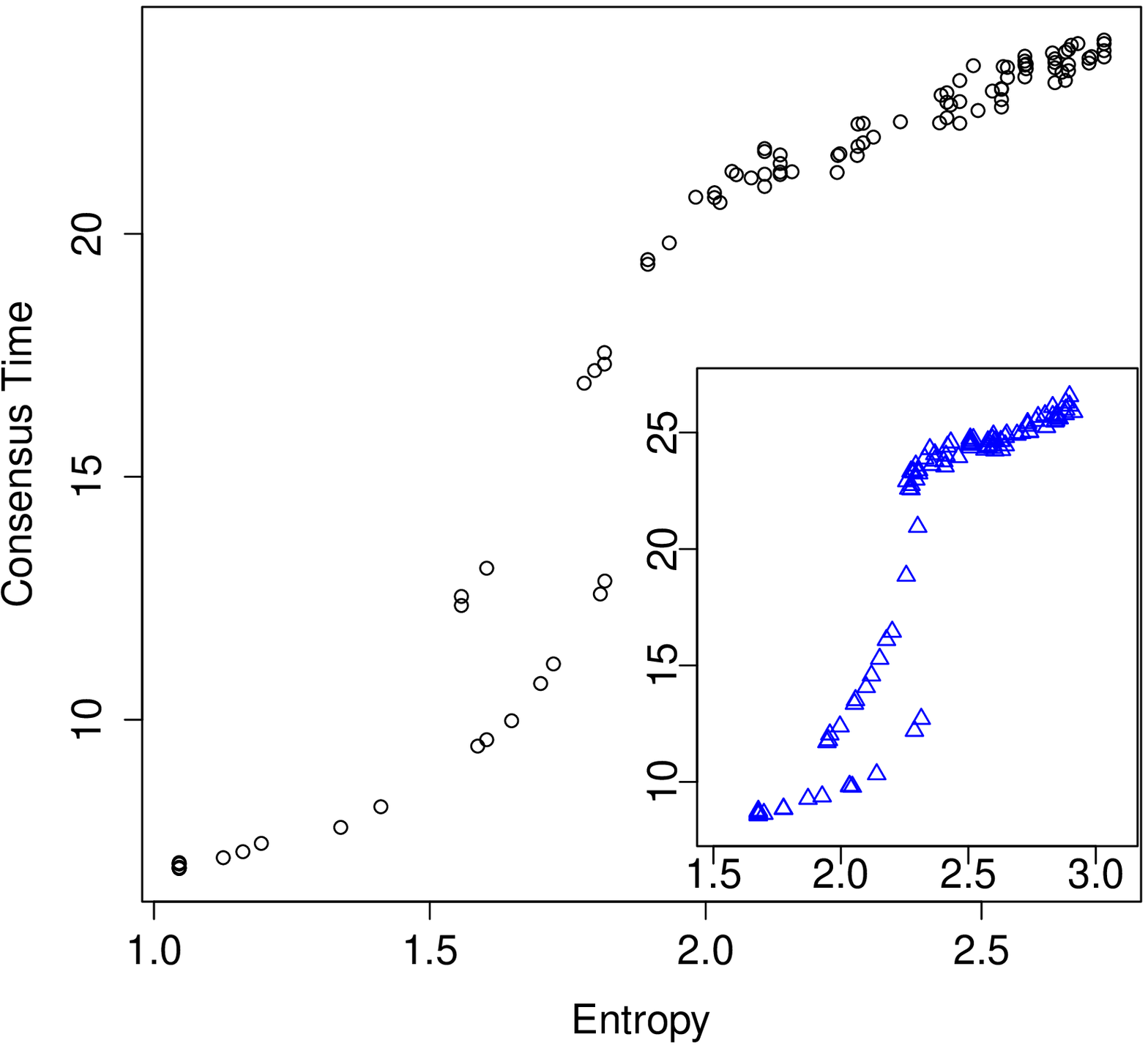}
\includegraphics[scale=.4]{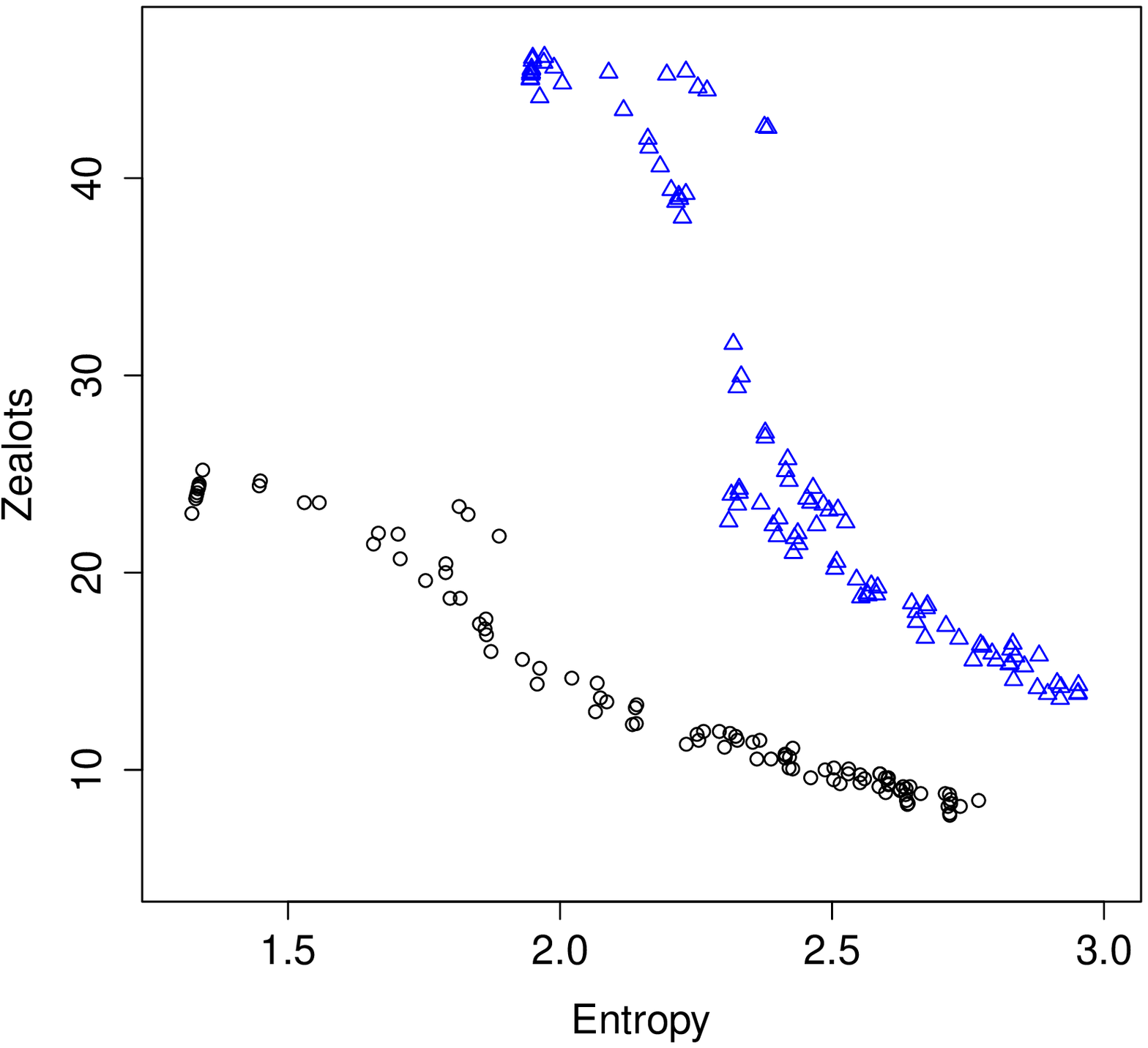}
\caption{Plots of the consensus time (top) and the critical number of zealots (bottom) against the Shannon entropy of various initial conditions. A committed minority of word $A_1$ is introduced only in the bottom figure. Data are shown for $N=200$, $K=20$ $(\circ)$ and $N=400$, $K=40$ $(\triangle)$.}
\label{fig1}
\end{figure}

Intuitively, the first item of the principle means that if there is more uncertainty and disagreement in the system, the more time it takes to reach agreement. In the case of the voter model with two opinions, the consensus time on the complete graph is exactly equal to this measure of entropy scaled by $N$ \cite{sood,starnini}. The posititve correlation between entropy and consensus time in the naming game is demonstrated in Fig. \ref{fig1}. The second item of the entropy principle suggests that if there is greater dissent in a population, then it is easier for a minority of zealots to dominate the system. This reinforces the \textit{if divided then conquered} maxim since it is easier to dominate the system in the presence of greater internal conflict. Figure \ref{fig1} demonstrates the effect of the entropy of the initial condition on the critical number of zealots.

\section{Rate of convergence}\label{sec:convergence}
Our analysis of the naming game is based on the rate of convergence of the system. The rate of convergence is given by the dominant eigenvalues of the transition matrix for the probability distribution of the system. Knowing the rate of convergence, we can estimate the time until a word is eliminated (collapse time) as well as the consensus time. For the case with committed minorities, we can also use this analysis to estimate the number of zealots required until a drastic qualitative change occurs in the system. This is because, the dominant eigenvalues of the transition matrix depend on the number of zealots. When the fraction of committed minorities is high enough, these eigenvalues no longer dominate the ordering of the system. This means that different eigenvectors determine the overall shape of the probability distribution over time and there is a significant change in qualitative behavior. Once we have the dominant eigenvalues, these solutions become easy to find.

To find the convergence rate, we first express the transition matrix componentwise. Let $ n_W(m)$ be the total number of individuals with word list $W$ at discrete time $m$, and let the vector $\bm n$ take components $n_W$. Also let 
\begin{equation}
a_{\bm\alpha}^{(m)}=Pr\{\bm n(m)=\bm\alpha\}.
\end{equation}
We seek to express $a_{\bm\alpha}^{(m+1)}$ in terms of $a_{\bm\alpha}^{(m)}$. To do this, we must account for all possible transitions that the model allows. Although this is a complicated task for the general $K$ word naming game, we follow a simplified model to ameliorate this issue while keeping the original qualitative properties intact. In the simplified model, only the listener updates their word list in response to a message from the speaker, as in \cite{marvel}. This we call the listener-only naming game. In every simulation, we apply the original naming game rules, which shows that there is still agreement under this modification.

Since we assume that only one individual changes their word list in a given time step, an individual with word list $W$ may transition to having word list $W'$ or vice versa. To account for all transitions in the stochastic matrix, we must consider all pairs of word lists $(W_1,W_2)$ along with their respective transition probabilities. Let $D$ be the set of all pairs of word lists. Also let $\mathcal{L}_I[\cdot]$ be the operator acting on the current macrostate that accounts for the possible transitions involving word pair $I=(W_1,W_2)$. We then write

\begin{equation}
a_{\bm\alpha}^{(m+1)}-a_{\bm\alpha}^{(m)}=\sum_{I\in D}\mathcal{L}_I\left[a_{\bm\alpha}^{(m)}\right].
\end{equation}

We estimate the rate of convergence of the model by the spectral properties of each $\mathcal{L}_I$. Summing all of them together gives the relative magnitude of $a_{\bm\alpha}^{(m+1)}-a_{\bm\alpha}^{(m)}$, which is the change in probability over a single time step. We wish to find the smallest change in probability possible that retains $K$ words in the system. Since each $\mathcal{L}_I$ corresponds to pairs of word lists that transition to each other, we exhaust each case of pairs of word lists and find the smallest eigenvalues, many of which are zero. The meaning of each case is that we only allow the given pair of word lists, $(W_1,W_2)$ to change their word lists in a given step.

\subsection{Case 1: $|W_1|,|W_2|\geq 2$}\label{sec:A}

These cases tend to a stationary distribution that is not the consensus state. If we only allow a pair of word lists that contain multiple words, then it is impossible to update the system in such a way that a word is eliminated. The only way for a word to be eliminated is if a listener is the only holder of it and hears and then adopts a familiar word. Since neither $W_1$ nor $W_2$ fits this criterion, we take the change to be $0$ without loss of generality. Note also that this conclusion applies to the vast majority of cases for large $K$.

If the system does not converge to consensus, then it converges to the stationary distribution acquired from these cases. It is valuable to understand the behavior of the second largest eigenvalues in these cases, especially when considering zealots. The rate of convergence to the stationary distribution yields the criteria for the phase transition as different sets of eigenvectors starts governing the shape of the system. The stationary distribution in this case is related to the metastable distribution when the number of zealots is small. So, we seek to find the size of the rate of convergence to this stationary distribution.

The only possible means of transition in this case occurs when $W_1$ and $W_2$ differ by a single word. Otherwise it is impossible for $W_1$ and $W_2$ to transition to each other. Let $W_2=W_1\cup \{A_p\}$ and let $S_p$ be the set of word lists that contain $A_p$. Note that there are $K$ different choices for $A_p$. Let $p_i$ be the probability of transition from $W_1$ to $W_2$ given that $n_{W_1}=i$, which is given by
\begin{equation}
p_i=\left(\sum_{W\in S_p}\frac{n_W}{N|W|}\right)\frac{i}{N-1}.\label{eigen_A}
\end{equation}
Since it is impossible to transition from $W_2$ to $W_1$ in the naming game, this constitutes a triangular transition matrix, whose spectrum is $\lambda_k=-p_k$. Let 
\begin{equation}
\mu_p=\sum_{W\in S_p}\frac{n_W}{N|W|},
\end{equation}
which depends on the macrostate of the system and the particular word pair. The total change in probability comes from the sum of the relative changes for each word. That is, we sum Eq. \eqref{eigen_A} for $p=1\ldots K$. In doing so, we find that the sum of $\mu_p$ is at most $O(1)$ if the sum of $n_W$ achieves its maximum value of $O(N)$. This yields a total rate of change being proportional to $1/N$ to leading order.

We are also interested in a second term in total change in probability, as it is significant for the naming game with zealots. This is attained by supposing that the sum over $\mu_p$ does not achieve its maximum value. If each $n_W$ is only $O(1)$, then the sum of $\mu_p$ is $O(K/N^2)$. This matches the leading term for $K=O(N)$, but is smaller for $K=O(1)$. These considerations are utilized when calculating the total rate of convergence.

\subsection{Case 2: $W_1=\{A_k\},|W_2|>2, A_k\in W_2$}\label{sec:B}

Here we only consider transitions in a word list that contains a single word and a word list that have three or more words. The size of the eigenvalues are easy to find in this case since it is only possible for $W_2$ to become $W_1$. This is because it is impossible for an individual with only a single word to adopt three or more words in a single step. Mathematically, this case corresponds to a triangular transition matrix, whose eigenvalues are the diagonal elements. Let $p_i$ be the probability that an individual with word $W_2$ hears word $A_k$ and thus transitions to $W_1$ given that there are $i$ individuals with $W_1$. Since all other individuals with all other word lists are considered fixed, let
\begin{equation}
\mu_1=\sum_{W\in S_k\backslash\{W_1\cup W_2\}} \frac{n_W}{|W|}\label{def:mu1},
\end{equation}
which is considered constant. Now, we express the transition probability as
\begin{multline}
p_i=\frac{(i+n')(N'-i)}{N(N-1)}+\frac{1}{|W_2|}\frac{N'-i}{N}\frac{N'-i-1}{N-1}\\
+\mu_1\frac{(N'-i)}{N(N-1)},
\end{multline}  
where $N'=n_{W_1}+n_{W_2}$, which is conserved here. Also, $n'$ is the number of zealots corresponding to the word $A_k$. The eigenvalues for this case are $-p_i$, and the smallest eigenvalue that does not correspond to consensus is 
\begin{equation}
\lambda\sim -\frac{N'+\mu_1+n'}{N^2}\label{case_B_rate}
\end{equation}
This can be seen by taking $i=N'-1$. Note that $\mu_1$ and $N'$ captures the dependence on the state of the system on the relative change in probability.

\subsection{Case 3: $W_1=\{A_k\},W_2=\{A_k,A_l\}$}\label{sec:C}

Here $W_1$ has only one word and $W_2$ has two words, one of which is $A_k$ for some $k$. This is the most dynamic of the cases because $W_1$ can transition to $W_2$ and vice versa. Because of the listener only assumption, this constitutes a tridiagonal transition matrix. Let $p_i$ and $q_i$ be the probability $n_{W_1}$ increases and decreases respectively, given that $n_{W_1}=i$. Let
\begin{equation}
\mu_2=\sum_{W\in S_l\backslash\{W_2\}} \frac{n_W}{|W|}
\end{equation}
and recall the definition of $\mu_1$ from Eq. \eqref{def:mu1}. The transition probabilities are then expressed as
\begin{align}
p_i&=\frac{(i+n')(N'-i)}{N(N-1)}+\frac{(N'-i)(N'-i-1)}{2N(N-1)}\\
&+\mu_1\frac{N'-i}{N(N-1)}\nonumber,\\
q_i&=\frac{i(N'-i)}{2N(N-1)}+\mu_2\frac{i}{N(N-1)}.
\end{align}
To find the rate of convergence for this step, we wish to solve the following eigenvalue problem
\begin{equation}
\lambda c_i=p_{i-1}c_{i-1}+(-p_i-q_i)c_i+q_{i+1}c_{i+1}\label{eigenvalue_problem}
\end{equation}
In order to solve for all eigenvalues of this problem, we apply the generating function method of Ref. \cite{pickering}, which exactly diagonalized the voter model. We begin by expressing Eq. \eqref{eigenvalue_problem} in terms of a generating function $G(x,y)$, which we define as
\begin{equation}
G(x,y)=\sum_{i=0}^{N'} c_i x^iy^{N'-i}
\end{equation}
Using shift and differentiation properties of $G$, we rewrite Eq. \eqref{eigenvalue_problem} as
\begin{multline}
N(N-1)\lambda G=(x-\frac{1}{2}y)(x-y)G_{xy}+(n'+\mu_1)(x-y)G_y\\
+\frac{1}{2}y(x-y)G_{yy}-\mu_2(x-y)G_x
\end{multline}
We solve this by the change of variables $u=x-y$ and $G(x,y)=H(u,y)$. Here, we have
\begin{equation}
H(u,y)=\sum_{i=0}^{N'} b_i u^iy^{N'-i}.\label{H}
\end{equation}
Making this change gives the equivalent equation for $H$:
\begin{multline}
N(N-1)\lambda H=\left(u^2-\frac{1}{2}uy\right)H_{uy}-u^2H_{uu}+\frac{1}{2}uyH_{yy}\\
+(n'+\mu_1)uH_y-(n'+\mu_1+\mu_2)uH_u.
\end{multline}
The above written as a difference equation for the coefficients of $H$ gives
\begin{multline}
N(N-1)\lambda b_i\\
=-\left[\frac{1}{2}i(N'-i)+i(i-1)+i(n'+\mu_1+\mu_2)\right]b_i\\
+(N'-i+1)\left[\frac{1}{2}N'+\frac{1}{2}i-1+n'+\mu_1\right]b_{i-1}.\label{b_recursion}
\end{multline}
This constitutes a lower triangular matrix problem for $b_i$. If there is not a singularity in $b_i$ for some $i$ between $0$ and $N'$, then all $b_i=0$, which is trivial. So, assuming that there exists a singularity at some $i=k$, we require the $b_i$ to vanish. This yields the following result for the eigenvalues of this case:
\begin{equation} 
\lambda_k=-\frac{k(k-1)+\frac{1}{2}k(N'-k)+(n'+\mu_1+\mu_2)k}{N(N-1)}.\label{eigenvalue_case3}
\end{equation}
Note that this result depends on the number of committed agents, $n'$. Each $b_i$ can be found explicitly by Eq. \eqref{b_recursion} by taking $b_k=1$ and $b_i=0$ for $i<k$. We then find the coefficient of $G(x,y)$ by calculating $H(x-y,y)$. Doing so gives
\begin{equation}
G(x,y)=\sum_{i=0}^{N'}\left[\sum_{j=i}^{N'}{j\choose i}(-1)^{j-i}b_j \right]x^iy^{N'-i}.\label{G_exact}
\end{equation}

The value of $c_i$ in terms of $b_j$ is given in the bracket of Eq. \eqref{G_exact}. To find the dominant eigenvalue of this case, take $k=1$ in Eq. \eqref{eigenvalue_case3}, which yields
\begin{equation}
\lambda=-\frac{\frac{1}{2}N'+n'+\mu_1+\mu_2}{N^2}.\label{case_C_rate}
\end{equation}
Similar to Eq. \eqref{case_B_rate}, the change in probability depends on the state of the system.

\subsection{Total rate of convergence}
Now that we have results for each case, we put them together to obtain the convergence rate of the naming game. We will make some assumptions about the state of the system. First, we assume that initially there is symmetry in the representation of words. That is, no word initially dominates the other words in accord to the applications given here. Second, we assume that for each word, there are individuals with only this word in their lists. The system quickly orders itself this way as long word lists are replaced by lists of length $1$. This second assumption allows us to utilize Cases $2$ and $3$ above when determining the rate of convergence.

The rate of convergence is estimated by the smallest non-zero change given by the above cases for $\mathcal{L_I}$. So, the rate of change of the probability distribution for a \textit{single word} $A_k$ is
\begin{equation}
1-\lambda_k=O\left(\frac{\theta+n'}{N^2}\right)\label{one_word_rate},
\end{equation} 
where $\theta=N'+\mu_1+\mu_2$, which describes the macrostate of the system. If we take this to be the total change in probability, then we have implicitly assumed that there are only two words in the system, and all others have been eliminated. So we require that all $K$ words are present in the system and sum the smallest change in probability given by Eq. \eqref{one_word_rate} for each word. By symmetry, the total change in probability is $K$ multiplied by the right hand side of \eqref{one_word_rate}. Therefore, the total rate of convergence is given by
\begin{equation}
1-\lambda=O\left[\frac{K(\theta+n')}{N^2}\right].\label{eigenvalue}
\end{equation}
We make use of Eq. \eqref{eigenvalue} extensively to determine the collapse time, consensus time, and the location of a phase transition over the number of zealots. We need to carefully account for the macrostate of the system when applying Eq. \eqref{eigenvalue} due to the presence of $\theta$. We expect the macrostate of the system to significantly affect the solution for the consensus time and phase transition. 

Now we wish to find the rate of convergence to the metastable state in the presence of committed minorities. These are given by case $1$ above. The largest of these was found to be $O(1/N)$ and the next largest was $O(K/N^2)$. Since the rate of convergence is given by the sum of these cases, we find that the rate of convergence to the metastable state is
\begin{equation}
1-\lambda\sim\frac{a}{N}+\frac{bK}{N^2}.\label{total_eigen_A}
\end{equation}
Here $a$ and $b$ are constants. When the convergence rate to the metastable state exceeds the convergence rate to consensus, the system is trapped in the metastable state. Otherwise, the system rapidly moves to consensus. This gives the criterion for the phase transition over $n'$.

\section{Naming game without zealots}\label{sec:no_zealots}
We start with the simple case when the system does not have committed minorities. That is, we take $n'_k=0$ for every word. Also, we assume that each word has near equal representation in the initial condition. That is, we do not assume that any given word significantly dominates any other word in the population. We also assume that none of the individuals have mixed word lists initially. Now $\theta$ can be as large as $O(N)$ since there can be $O(N)$ individuals with words lists of length $2$ or more. Even though $\theta=O(1)$ initially, the system quickly saturates itself with individuals that have longer word lists. Assuming that this is the case, we take $\theta=O(N)$ in Eq. \eqref{eigenvalue} and have
\begin{equation}
\lambda\sim 1-O\left(\frac{K}{N}\right)\label{no_zealot_rate}
\end{equation} 
as the rate of convergence. We apply this to calculate the amount of time until an opinion is eliminated from the system entirely, which allows us to estimate the number of states over time as well as the consensus time.

\subsection{Collapse time}
We define the collapse time as the amount of scaled time until a word is eliminated from the system. Scaled time is the number of discrete time steps divided by the number of nodes in the network. That is, the scaled time $t$ is defined by $t=m/N$. Now we wish to find the amount of time until the system is expected to transition from having $k$ words to at most $k-1$ words. 

If the system is not near consensus, then it is not dominated by the diffusion terms in the random walk. That is, the entire probability distribution cannot be estimated above by the dominant eigenvalue when away from consensus. When this is the case, we take the survival probability, which is the probability that there are $k$ words in the system at time $t$, and set it to $1/N$. When this is the case, it is expected that less than one individual will have one of the $K$ words. Given that there are $k\leq K$ words in the system at scaled time $t$, the survival probability is $\lambda^{tN}$. So, using this criterion, we have that
\begin{equation}
\lambda^{\tau_kN}=\frac{1}{N}
\end{equation}
which implies that 
\begin{equation}
\tau_k^{(outer)}=O\left(\frac{\ln N}{k}\right).\label{def:outer}
\end{equation}
We use the notation $\tau_k^{(outer)}$ to designate that this holds when the system is not near consensus. When the system is near consensus, the system is diffusionlike, we use the infinite series to calculate the expected value. That is, the collapse time near consensus, $\tau_k^{(inner)}$, is given by
\begin{equation}
\tau_k^{(inner)}=\sum_{m=0}^\infty s_m \frac{m}{N}
\end{equation} 
where $s_m$ is the probability of collapse. The probability of collapse is the change in the survival probabilities: $s_m=\lambda^{m-1}-\lambda^m$. Making this substitution into the infinite series for $\tau_k^{(inner)}$ gives
\begin{equation}
\tau_k^{(inner)}=O\left(\frac{1}{k}\right).\label{def:inner}
\end{equation}
The collapse time from the outer region differs from the collapse time from the inner region by a factor of $\ln N$. As the system approaches consensus, this $\ln N$ tends to $O(1)$ as the system transitions from one region to another. We make use of these observations, as well as the collapse times in the following section.

\subsection{Opinions over time and Consensus time}\label{sec:states_consensus}
\begin{figure}[h!]
\includegraphics[scale=0.5]{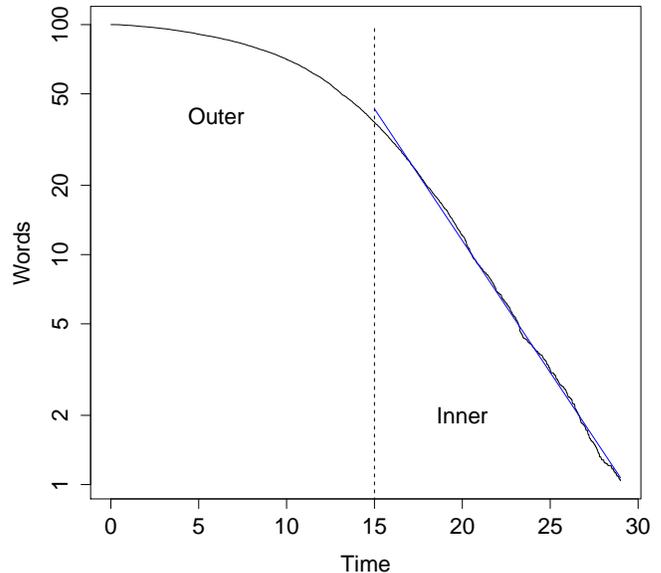}
\caption{Semi-log plot of the number of words in the system as a function of scaled time $t$. Notice that it takes longer for words to be eliminated when the system is in the outer region. Also shown is a line of best fit for the inner region, which confirms that the states over time tends to an exponential in $t$. The naming game is averaged over $100$ runs with $N=K=100$.}\label{fig2}
\end{figure}

Here we estimate the number of words in the system over time as well as the time to consensus. To estimate the number of words over time, we sum the collapse times for their respective regions to find these quantities. Starting with the outer region, we estimate the time it takes to achieve $S$ words by
\begin{align}
t&=\sum_{k=S}^{K}O\left(\frac{\ln N}{k}\right)\\
&=O\left(\ln N \ln\frac{K}{S}\right).\label{outer_time}
\end{align}
Solving for $S$ gives
\begin{equation}
S_{outer}(t)\leq K\exp\left(-\frac{\alpha t}{\ln N}\right).
\end{equation}
This shows an exponential convergence in $S$ on a logarithmic scale for $t$. Repeating this process for the inner region shows the time to reach $S$ words is
\begin{equation}
t=O\left(\ln \frac{K}{S}\right)\label{inner_time},
\end{equation}
Solving for $S$ gives
\begin{equation}
S_{inner}(t)\leq K\exp(-\alpha t).
\end{equation}
The inner region converges on a faster time scale than the outer region. The convergence of the outer region will, however, accelerate as the system approaches consensus. These results are shown numerically in Fig. \ref{fig2}.

Now we will estimate the consensus time, which is defined as the total amount of scaled time until the entire system adopts a single word. For the naming game with two words, $A$ and $B$, the consensus time is $O(\ln N)$ \cite{castello,baronchelli2,baronchelli3}. However, this may vary when the number of words is large. Also of importance is the fact that each word is equally represented to acquire an upper bound on the consensus time. This is due to the observation that the consensus time increases with entropy. 

\begin{figure}[h!]
\includegraphics[scale=0.5]{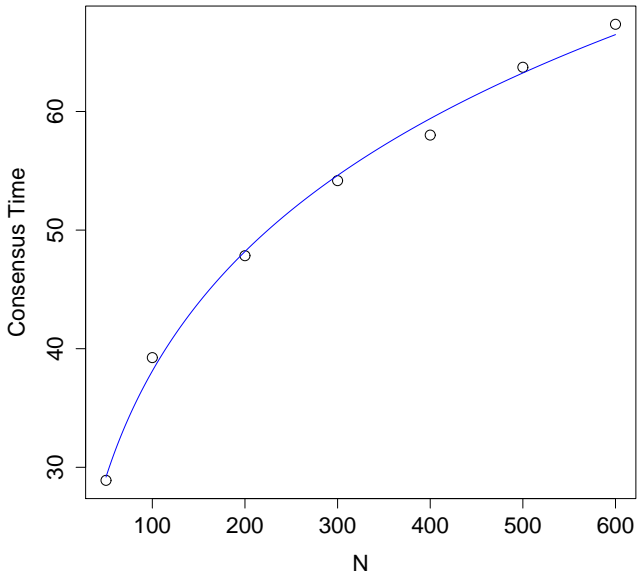}
\caption{Plot of the consensus time averaged over $50$ runs of the naming game for various $N$ with $K=N$. Also plotted is the estimate given in Eq. \eqref{consensus_time} fitted to the data. The best fit yields $c_1=1.18$ and $c_2=2.84$ in the context of Eq. \eqref{consensus_time}. The figure shows good agreement with the theory of Sec. \ref{sec:states_consensus}.}\label{fig3}
\end{figure}

To find the consensus time, we estimate the time spent in the outer and inner regions. Once this is known, the time to consensus is given by the sum of these two. To do this, we make use of Eqs. \eqref{outer_time} and \eqref{inner_time}. Since we do not know exactly which value of $S$ is the transition point when the system is near consensus, we take $S=1$ to yield an upper bound for the consensus time. With these assumptions, we have
\begin{align}
t_{outer}&=O(\ln N \ln K)\\
t_{inner}&=O(\ln K)
\end{align}
Therefore, the expected time to consensus is
\begin{equation}
E[\tau]\sim c_1\ln N \ln K+c_2\ln K\label{consensus_time}
\end{equation}
where $c_1$ and $c_2$ are constants. This is consistent with known information regarding the case when $K=2$, for Eq. \eqref{consensus_time} is $O(\ln N)$ for $K=2$. It also accounts for cases when $K$ takes extreme values. For $K=O(N)$, the consensus time increases to $O(\ln^2 N)$. An example of an extreme $K$ case is given in Fig. \ref{fig3}.

\section{Naming game with zealots}\label{sec:zealots}
Here we consider the case in which there are zealots in the system. If a zealot hears an unfamiliar word, then the new word is not added to the zealot's word list. We will use the above theory on the rates of convergence to analyze the properties of the naming game when zealots are included. We consider two cases. The first case is when the zealots all have the same word. The second case is when there are an equal number of zealots with each word.

\subsection{Zealots of one word}
This system assumes that all zealots share the same word. Without loss in generality, let us say that there are $n'$ zealots with word $A_1$. We do not assume here that there are any zealots with words $A_2,\ldots,A_K$. In the case where $K=2$, a bifurcation occurs over the number of zealots. It has been shown that when $n'/N\approx 10\%$, there are enough zealots to quickly turn an entire population. If the fraction of zealots is below this value, then the system is trapped in a metastable state, and it takes an exponential time for the population to adopt the zealots' opinion \cite{xie,waagen}.

We seek to extend this to cases when $K$ is arbitrary. Particularly, we consider cases when $K$ is large and the spectral method is required to analyze the system. This problem was briefly discussed in Waagen \textit{et al.} \cite{waagen} and their conclusion was that the same $10\%$ critical fraction holds for all $K$ and initial conditions to guarantee the zealots dominate the system. Their approach is to consider the worst case initial conditions and show it reduced to the $K=2$ case. The worst case initial condition minimizes entropy, and according the the entropy principle above, this maximizes the number of zealots required. We take the analysis of Waagen \textit{et al.} \cite{waagen} a step further by assuming the opposite scenario for the initial condition: Each uncommitted community is initially of equal size, which maximizes entropy.

Let $C$ be the number of individuals initially with word $A_k$, where $A_k$ is not the zealots' opinion. For the case when there are only zealots of a single type, we have $N=(K-1)C+n'$. For fixed $N$, this gives a dependence on $C$ in terms of $K$, given by
\begin{equation}
N=(K-1)C+n'.\label{N_onetype}
\end{equation}
Of particular interest is the dependence of the critical number of zealots, $n'_c$, on $K$, $N$, and $C$. By Eq. \eqref{N_onetype}, if we keep $N$ fixed, then the dependence on $C$ can be found from the dependence on $K$ by substitution.

\begin{figure}[h!]
\includegraphics[scale=0.4]{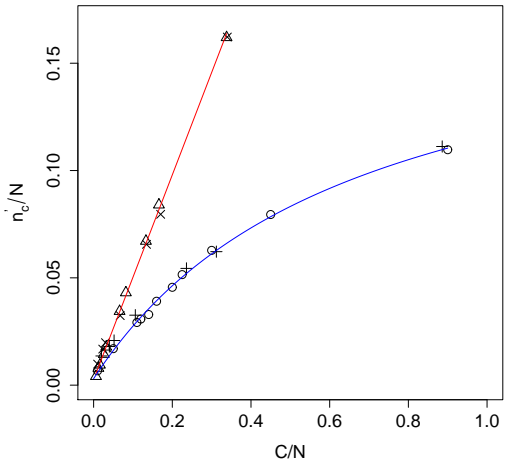}
\caption{Critical fraction of committed agents plotted against $C/N$ for $N=1000$ $(\circ, \triangle)$ and $N=500$ $(+,\times)$. Data for a committed minority of a single word is shown $(\circ,+)$ and the best fit of Eq. \eqref{nc_v_C2} in blue. Also shown is the case when there are committed minorities of every word $(\triangle,\times)$ with the fit of Eq. \eqref{all_zealots} in red.}\label{fig4}
\end{figure}

To find the phase transition over $n'$, the criterion we use is simple. This occurs when Eq. \eqref{eigenvalue} is dominated by a different class of eigenvalues that describe a stationary distribution. This stationary distribution is the metastable state, and the system will converge to it if the rate is higher than the consensus rate. The rate to the metastable state is given in Eq. \eqref{total_eigen_A}. Setting Eq. \eqref{eigenvalue} equal to Eq. \eqref{total_eigen_A} gives 
\begin{equation}
1-\lambda=\frac{a}{N}+\frac{bK}{N^2},\label{criterion}
\end{equation}
where $a$ and $b$ are constants. We take $\theta=O(1)$ in $1-\lambda$ since the system is initially dominated by uncommitted words. Taking Eq. \eqref{criterion} and solving for $n'$ gives
\begin{equation}
n'_c=\frac{aN}{K}+b.\label{nc_v_K}
\end{equation}
Here $a,b=O(1)$. This tells us that we expect the number of zealots required to turn a population decays as $1/K$ to a constant. We express this in terms of $C$ by substitution. This produces a nonlinearity in $n'_c$, which we approximate to provide the following fit:
\begin{equation}
n_c'=\frac{aNC}{N+dC}+b,\label{nc_v_C2}
\end{equation}
where $d$ is another constant. A comparison of this against simulation data is given in Fig \ref{fig4}. This result shows that as the relative sizes of the community grows larger, it takes more zealots to turn the population.

\subsection{Zealots of every word}

The case where each opinion has zealots follows by a similar argument. If $n'< n'_c$, then one opinion eventually will suppress all others. When $n'>n'_c$, a stalemate develops and no opinion gains dominance. We still apply criterion of Eq. \eqref{criterion} along with Eq. \eqref{eigenvalue} for the phase transition. This means that the dependence of the critical number of zealots as a function of $K$ has the same form as Eq. \eqref{nc_v_K}. However, now we have $N=K(C+n')$. When substituting $K$ for $C$, we obtain
\begin{equation}
n_c'=a'C+b'.\label{all_zealots}
\end{equation}
We use $a'$ and $b'$ to denote different constants from the previous case that are also both $O(1)$. Fig. \ref{fig4} depicts this relationship in practice.

\section{Discussion}\label{sec:discussion}


The first contribution presented here is technical. We introduce an innovative approach to deal with a large number $K$ of opinions, that require analyzing $O(2^K)$ equations in the traditional ODE-based approach. Another contribution is advancing our understanding of naming game dynamic by considering  its dependence on the initial condition. We demonstrate that the consensus time and the critical number of zealots have distinct correlations with the entropy of the state. This reinforces the divide and conquer rule and also suggests that social systems with great dissent can foster many committed minority groups that may block each other from reaching a tipping point, which is high in case of the uncommitted groups sharing a few opinions only. Our results suggest that high opinion diversity among uncommitted individuals changes the dynamics. In such situations, the tipping point can be reached with the number of committed minority members being small, or even independent of the system size, making the system unstable and quickly transferring to the state in which uncommitted individuals adopt one of the minority opinions.

\begin{acknowledgments}
This work was supported in part by the Army Research Office Grants No. W911NF-09-1-0254 and No. W911NF-12-1-467 0546, the Army Research Laboratory under Cooperative Agreement No. W911NF-09-2-0053, the Office of Naval Research Grant No. N00014-15-1-2640, the European Commission under the 7th Framework Programme, Grant Agreement No. 316097 (ENGINE), and the National Science Centre, Poland, Decision No.DEC-2013/09/B/ST6/02317.We also thank G.Korniss for beneficial discussions. The views and conclusions contained in this document are those of the authors and should not be interpreted as representing the official policies, either expressed or implied, of the Army Research
Laboratory or the U.S. Government. The U.S. Government is authorized to reproduce and distribute reprints for Government purposes notwithstanding any copyright notation thereon.
\end{acknowledgments}

\bibliography{KNG_PRE_revised}{}

\end{document}